\documentclass[12pt]{article}
\usepackage{latexsym}
\usepackage{graphics}
\usepackage{epsfig}
\usepackage{epstopdf}
\usepackage{amsmath}
\usepackage{cite}
\usepackage{hyperref}
\usepackage[caption=false]{subfig}

\usepackage{amssymb}

\newcommand{\be}{\begin{equation}}
\newcommand{\ee}{\end{equation}}
\newcommand{\beq}{\begin{eqnarray}}
\newcommand{\eeq}{\end{eqnarray}}
\newcommand{\beqn}{\begin{eqnarray*}}
\newcommand{\eeqn}{\end{eqnarray*}}
\newcommand{\ba}{\hspace*{-5pt}\begin{array}}
\newcommand{\ea}{\end{array}}

\newcommand{\bit}{\begin{itemize}}
\newcommand{\eit}{\end{itemize}}
\newcommand{\ben}{\begin{enumerate}}
\newcommand{\een}{\end{enumerate}}

\begin{document}

\begin{center}
\textbf{\Large Solitary wave dynamics governed by the modified FitzHugh-Nagumo equation} \vspace{0.5 cm}

 Aleksandra Gawlik$^{a,}$\footnote{e-mail:
\url{aleksandramalgorzatagawlik@gmail.com}}, Sergii  Skurativskyi$^{b,}$\footnote{e-mail: \url{
skurserg@gmail.com}}, Vsevolod  Vladimirov$^{a,}$\footnote{e-mail:
\url{vladimir@mat.agh.edu.pl}} \vspace{0.5 cm}

{\small \it
$^a$ Faculty of Applied Mathematics, \\
AGH University of Science and Technology, \\
Mickiewicz Avenue 30,  30-059 Krak\'{o}w, Poland,

$^b$ Division of Geodynamics of Explosion, \\
 Subbotin Institute of Geophysics, NAS of Ukraine,  \\
Acad. Palladina Avenue 32,  03142 Kyiv, Ukraine}


%
\end{center}

\begin{quote} \textbf{Abstract. }{\small 
The paper deals with the studies of the nonlinear wave solutions supported by the modified FitzHugh-Nagumo (mFHN) system.
It was proved in our previous work that the model, under certain conditions, possesses a set of soliton-like traveling wave (TW) solutions. In this paper we show that the model has two solutions of the soliton type differing in propagation velocity. Their location in parametric space, and stability properties  are considered in more details.  Numerical results accompanied by the application of the Evans function technique prove the  stability of fast solitary waves and instability of slow ones. A possible way of formation and  annihilation of localized regimes in question is studied therein too. 
}
\end{quote}

\begin{quote} \textbf{Keyword: }{\small 
modified FitzHugh-Nagumo model; solitary wave solutions; spectral stability; soliton collision; the Evans function technique; birth and annihilation of solitary waves
}
\end{quote}

\vspace{0.5 cm}

\section{Introduction}
The transport phenomena play a significant role in open dissipative systems, in which self-organization processes are most pronounced \cite{Makar97,DDMSV}. For this reason, the models of these phenomena are the subject of numerous studies both in the field of applied mathematics and in natural sciences. The comprehensive understanding of mechanism of transfer processes has been achieved up to date for relatively simple systems in which the deviations from the state of thermodynamical equilibrium are small. In this case  the linear approach is applicable,  leading  to the well-known wave and diffusive equations.

However, in living systems and other active  strongly  nonequilibrium media the transport processes manifest a different properties \cite{Dodd,KPP,Davydov,Kar03}. Among others,  the wave dynamics in nerve demonstrates the specific response creation when a stimulus threshold exceeds \cite{NSE, A_Scott}.  This response propagates in the form of solitary wave, behavior of which essentially differs from that of linear wave. 

 The first attempts to simulate this phenomenon have been performed in papers \cite{HodgHuxley,Nagumo1962}. A simplified form of this model (making it easier to analyze and use) was studied in  \cite{FitzHugh}. 
An even simpler model, making possible the obtaining exact solutions, was proposed in paper \cite{McKean,Wang} and investigated in more details in \cite{RinzK,Feroe_78}. Despite of considerable successes in the nerve dynamics modeling, there are still many problems concerning the justification of mathematical models  used. In other words, the starting models for pulse transmission belong to the class of parabolic equations leading to infinite velocities of wave propagation, which is unphysical. To agree the observed wave dynamics and properties of model applied, the hyperbolicity of equations can be incorporated into the model   from the very beginning. Using the improved axon model \cite{Engelbrecht1992} or the internal variables approach \cite{Maugin1994,V_Lik,Joseph,Kar03,Skur2019}, the following modified FitzHugh-Nagumo is proposed: 
\begin{equation}\label{PDEq}
\begin{array}{c}
\tau  v_{tt}+v_t=v_{xx}+f(v)-w,\\
w_t=\epsilon (v-\gamma w ),\\
\end{array}
\end{equation}
where $f(v)$ is the phenomenological source term, $\gamma >0$,  $ \epsilon >0$, and $ \tau >0$ is the  time of relaxation. Note that the model with $f(v)=v(v-a)(1-v)$ 
as well as the model with piecewise ($Z$-shaped) continuous function $f(v)$ is suitable for both the analytical  \cite{Hastings_76, Hastings_82,V_Lik, Arioli2015} and qualitative \cite{LusapaOM,LusapaRocky} treatment. 

 In case when $f(v)=v-V_R^2 v- V_R v^2-v^3/3\equiv v(v/\sqrt{3}-A)(Q-v/\sqrt{3})$, where $V_R$ controls in fact the coordinates of nontrivial roots, 
the model has been studied  by Miura \cite{Miura1982}. Both of these cubic functions are evidently related by means of the scaling transformation. Indeed,  the transformations
\begin{equation*}
\begin{split}
\tilde v=v/A\sqrt{3}, \tilde w=w/A^3\sqrt{3}, \tilde c=c/A^2, \tilde \gamma=\gamma A^2, \tilde \epsilon= \epsilon/A^4, \\
\tilde \tau=A^2(A^2+\tau c^2-1)/c^2, a=Q/A
\end{split}
\end{equation*}
allow one to pass from the system studied by Miura to the system (\ref{PDEq}) with $f(v)=v(v-a)(1-v).$

According to the theoretical studies performed in \cite{Gawlik2019},  the modeling system  (\ref{PDEq}) still supports  the solitary TW solutins, whose existence have been proved earlier for the case $\tau=0$ in \cite{Hastings_76,Hastings_82,Arioli2015,Carter2015,Carter2016}. However, the comprehensive  numerical experiments with the solitary wave evolution, creation and interaction  were not performed in \cite{Gawlik2019}. The aim of this work is to fill in this gap by elucidating the properties of solitary wave dynamics within the framework of the mFHN model (\ref{PDEq}).  

The structure of the paper is following. In Sec.\ref{Sec:one} we outline the results of phase space analysis for the dynamical system possessing   the homoclinic solutions  which relate to the  solitary wave profiles. The solitary wave location in the parametric space is considered therein too.  Sec. \ref{Sec:two} presents  the results of numerical studies of single solitary waves' evolution. The analysis of spectral stability of solitary waves, based on the Evans function technique, is performed numerically in Sec. \ref{Sec:three}.  The creation of fast solitary waves in piston problem  and the mutual annihilation of a pair of identical  fast solitary waves moving in opposite directions are discussed in Sec. \ref{Sec:four}. The paper ends up with concluding remarks at  Sec. \ref{Sec:concl}.

\section{Trapping the solitary wave solution to the mFHN equation}\label{Sec:one}

In this studies we use  Miura's version of the source function $f(v)$. 
We thus are going to treat  the traveling wave solutions  $v(t,\,x)=v(\xi),$ $w(t,\,x)=w(\xi),$ $ \xi=x+ct$, where $c$ is the wave velocity. 
Inserting this anzatz  into (\ref{PDEq}),  we obtain the following ODE system:
\[
\begin{array}{c}
\tau \,c^2\,v''= v'' - c v' +f(v)-w,\\
c w'= \epsilon (v-\gamma w ),
\end{array}
\]
Introducing the parameter  $\beta=\frac{1}{(1-\tau c^2)} >0 $ and  new variable $u=v'$, we finally get the dynamical system
\begin{equation}\label{VGS:uklad2}
\begin{array}{l}
v'=u, \\
u'=  \beta c  u - \beta f(v)+\beta w,\\
 w'= \frac{\epsilon}{c} (v-\gamma w ),
\end{array}
\end{equation}
describing the family of TW solutions.
Let us mention that due to  the condition $\beta\,>0\,$ the velocity $c$ cannot be arbitrarily large (which is unphysical) because of the inequality $c^2<1/\tau$. We also note that when replacing the inequality by $\beta\,<0\,$, neither in this model, nor in its simplified analogue \cite{V_Lik}, soliton solutions were not found and there is the reason to believe  that either they do not exist when $1/\tau<c^2,$ or they are unstable, and therefore unobservable.

As is was shown in \cite{Miura1982},  at $\tau=0$ the system (\ref{VGS:uklad2}) possesses the homoclinic orbit bi-asymptotic to the origin. Now we want to ensure that the homoclinic loop also remains in the modified equation when the  term with $\tau>0$ is added. 
\begin{figure}
\centering
\subfloat[]{
\resizebox*{6cm}{!}{\includegraphics{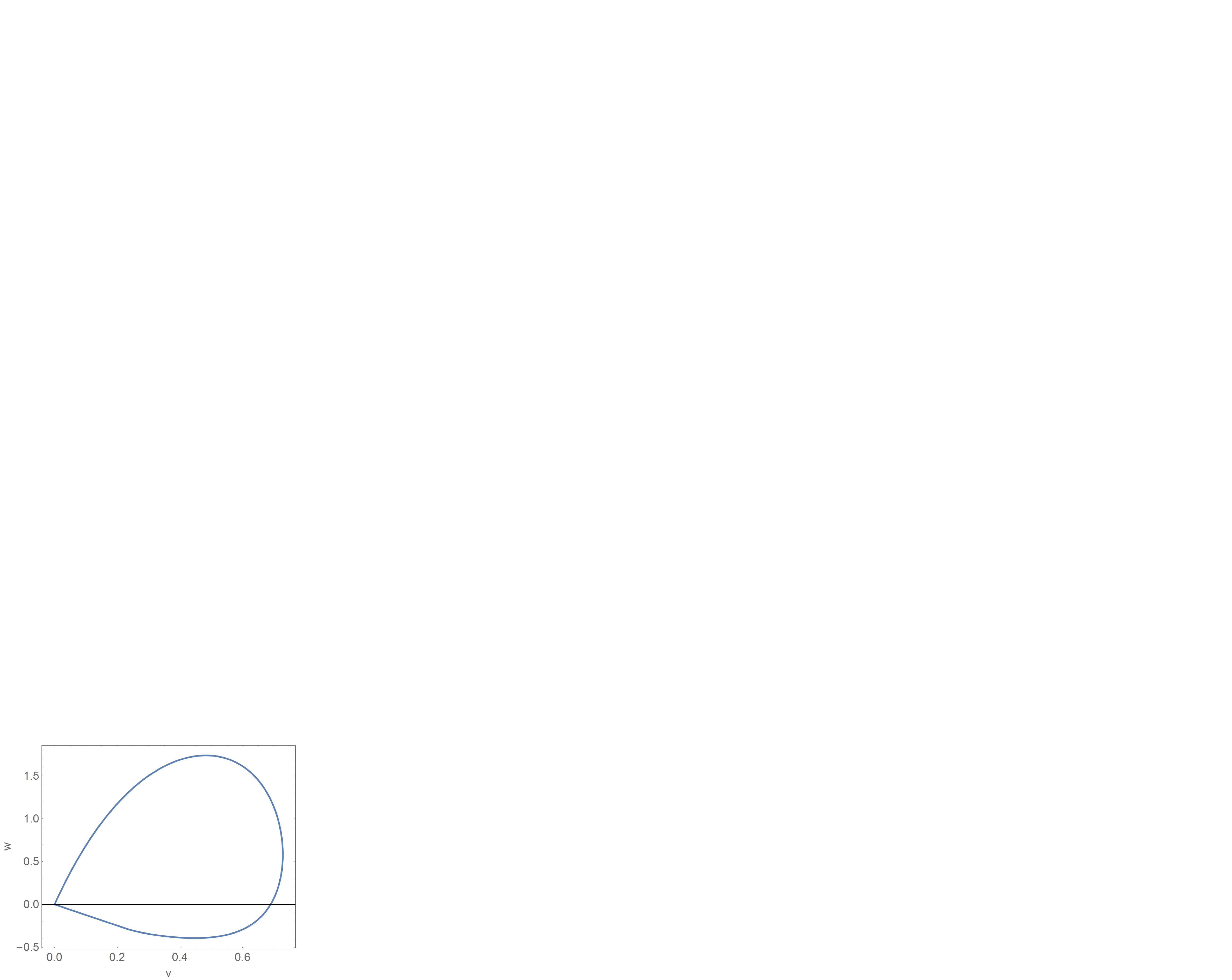}}}\hspace{5pt}
\subfloat[]{\resizebox*{6cm}{!}{\includegraphics{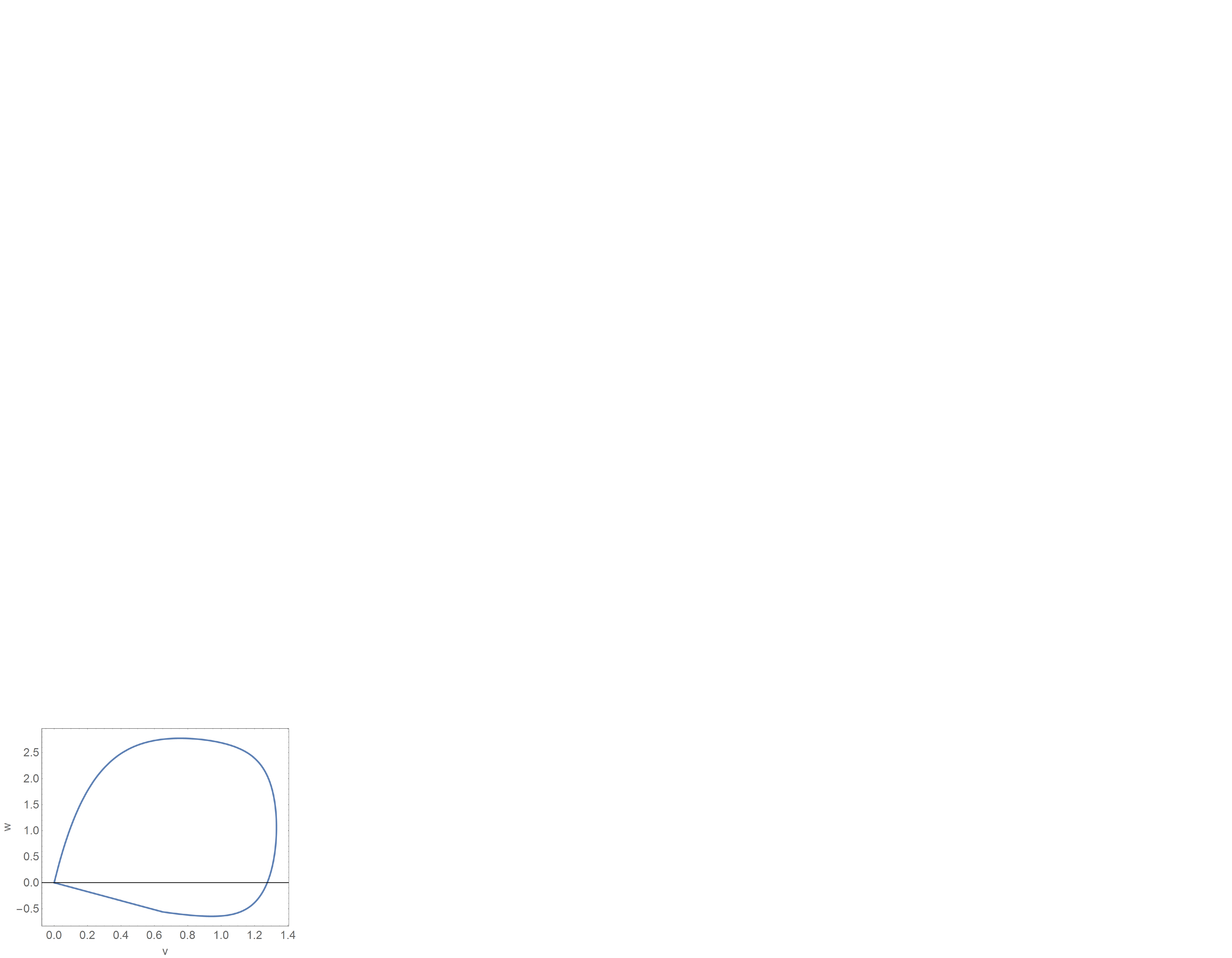}}}\hspace{5pt}
\subfloat[]{\resizebox*{6cm}{!}{\includegraphics{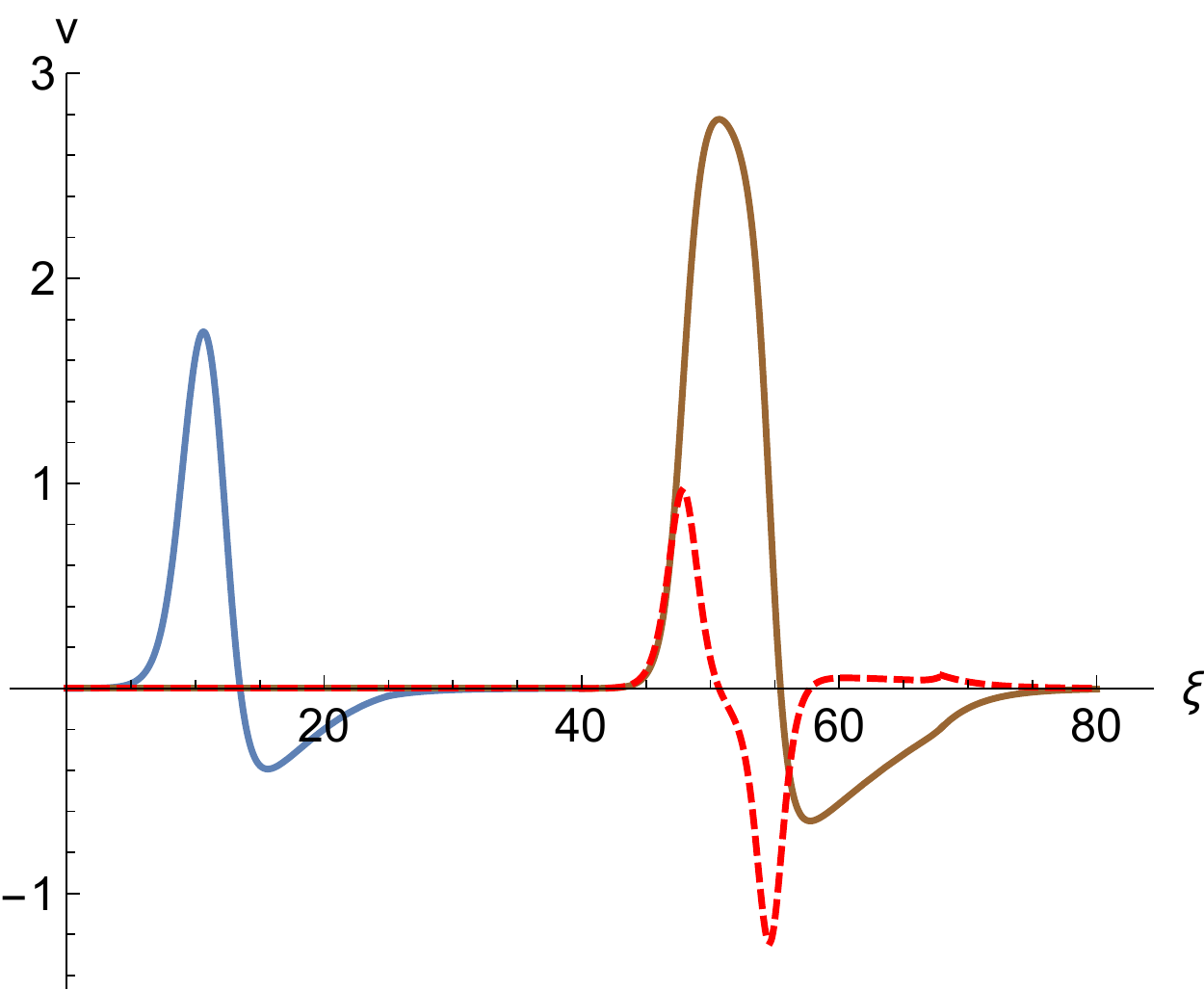}}}
\caption{Phase portraits of the homoclinic loops corresponding to the slow (a) and fast (b) solitary  waves, and their profiles (c) (solid line and  dashed line, correspondingly).  The parameter values are $\gamma=0.8$, $\epsilon=0.08$, and  $\tau=0.2$. } \label{vgs:fig1}
\end{figure}
We start building a homoclinic curve with an analysis of the linearization of system (2) at the origin:
\begin{equation}\label{VGS:linear}
\left(\begin{array}{c} v\\u\\w\end{array}\right)'=
\left(\begin{array}{ccc}   0  & 1 &  0 \\
 \beta (V_R^2-1) & \beta c  & \beta \\
 \epsilon c^{-1} & 0 & -\gamma \epsilon c^{-1} 
\end{array}\right)\left(\begin{array}{c} v\\u\\w\end{array}\right)\equiv J \left(\begin{array}{c} v\\u\\w\end{array}\right).
\end{equation}
The eigenvalues of the linearization matrix $J$ satisfy the characteristic equation
\begin{equation}\label{VGS:chareq}
W(\mu)=\frac{\beta \epsilon}{c}(1 + (V_R^2-1) \gamma ) + \beta ( V_R^2-1   + \epsilon \gamma  ) \mu +(c \beta -\frac{ \epsilon \gamma}{c} )\mu^2- \mu^3 =0.
\end{equation} 
We assume that the parameter's $V_R$ value is such that the inequality  $W(+\infty)W(0)<0$ is satisfied. Under this condition, the equation   (\ref{VGS:chareq}) possesses one positive real root, provided that the remaining  parameters are positive. This means that there exists a one-dimensional unstable manifold $W^u_{loc}$ starting from the origin. In fact, the unstable manifold has two branches, and we associate with the homoclinic trajectory that one, which is directed towards the first octant of the phase space. If the linearization matrix, in addition to the above mentioned positive eigenvalue, has a pair of roots with  negative real parts, then the assumed homoclinic trajectory will return to the origin along the two-dimensional unstable manifold $W^s_{loc}$, which is tangent to the linear space spanned by the corresponding pair of eigenvectors. These conditions allow us to state the parameter constrains providing the possibility for  homoclinic loop to exist. Advanced studies of the homoclinic orbit creation are based upon the small parameter technique application \cite{Gawlik2019}, which gives an additional information about the location of homoclinic trajectory in the parametric space. Taking this into account, we fix the parameters' values as follows:  
 $\gamma=0.8$ and $\epsilon=0.08$  (in accordance with Miura's suggestions  \cite{Miura1982}),  $V_R=-1.2$, $\tau=0.2$. In this case the eigenvalues of the matrix $J$ attain the values $\mu_{1,\,2,\,3}=\{1.308, -0.265 \pm 0.229 i\}$.  We start from the initial data $(u,v,w)=\nu \vec \ell_1$, $\nu=10^{-4}$, where $\vec \ell_1=(1,\mu_1,\epsilon/(\gamma \epsilon+\mu_1 c ))^{tr}$ is the eigenvector  corresponding to the eigenvalue $\mu_1>0$. Integration  of dynamical system (\ref{VGS:uklad2})
with the appropriate initial data by means of standard command from the ''Mathematica'' package allows us to identify the interval of velocities $[c_L;c_R]$, where $c_{L},\,c_{R}$ are the values of the parameter $c$ for which the  trajectory tends  to $- \infty$ and $+\infty$, respectively.

Using the command \textsf{ParametricNDSolve[$\dots$]}, the  value of $c^\star$ is chosen with high accuracy for which the orbit, leaving the stationary point along $W^u_{loc},$ approaches it at large values of $\xi$ as close as possible. 
Using this procedure, two homoclinic trajectories are identified at $c_1=0.526159710231696$ and $c_2=0.762091422672806$. The wave corresponding to the smaller value of $c$ is traditionally called the slow solitary wave, whereas that corresponding to larger one -- the fast wave. Their phase portraits are depicted in Figure~\ref{vgs:fig1}. Note that the fast wave  (Fig.~\ref{vgs:fig1}(b)) is higher and wider in comparison with the slow one (Fig.~\ref{vgs:fig1}(a)). Both wave profiles are unsymmetrical, which is connected with the dissipative character of the model.  

\begin{figure}
\centering
\resizebox*{7cm}{!}{\includegraphics{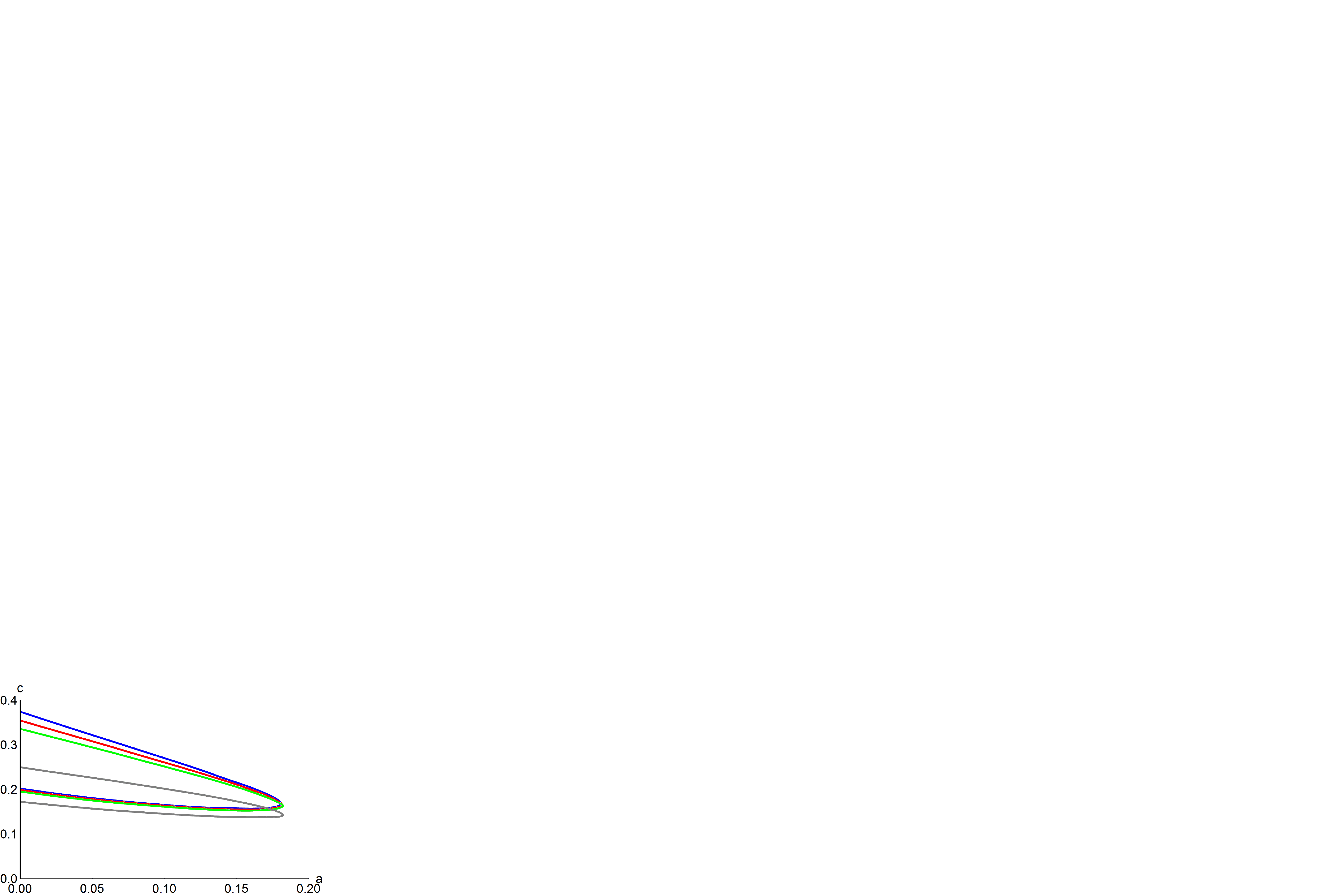}}
\caption{The location of fast and slow solitary waves in the parametric space $(a,\,c)$. 
} \label{vgs:fig2}
\end{figure}

No less interesting is the distribution of homoclinic orbits in the two dimensional parametric space  $(a,c)$, see Figure~\ref{vgs:fig2}. The curves  are plotted for $\tau=0.01, 0.1, 0.2, 1.0$ (the smaller the value of $\tau$, the higher is the corresponding curve in this figure).  Note that variation of the parameter $\tau$ does not change the shape of  the curve in essential way. All curves look like ''bananas''  observed at $\tau=0$ for the classical FHN model  \cite{Carter2016}. The upper branch of ``banana'' corresponds to the fast solitary waves, while  the lower one relates to the slow waves. However, a distinctive feature of the graphs obtained for our model is that the parameter $c$ does not rush to infinity when $a$ approaches zero from the right. When the parameter $\tau$ grows, the fast wave  velocity reduces more rapidly than that of slow wave, and the gap between them becomes smaller.  
We note in conclusion that the coordinates of the inflection point of the curves are almost independent of the value of the parameter $\tau$.

\section{Simulation of the single solitary wave movement}\label{Sec:two}

Consider now the evolution of the single solitary wave defined by the solution $Y_s=(v_s; u_s;  w_s)$ of the dynamical system (\ref{VGS:uklad2}). At first, let us chose the profile of slow solitary wave, see Figure~\ref{vgs:fig1}(a), as an initial condition for the PDE (\ref{PDEq}). In this case, $v(t=0,x)=v_s(\xi)$, $v_t(t=0,x)=c u_s(\xi)$, $w(t=0,x)=w_s(\xi)$, $x\in [0;L]$, the boundary conditions are periodic, i.e. $v(t,x=0)=v(t,x=L)$, $w(t,x=0)=w(t,x=L)$. Note that the initial profile needs some correction in the tail region, because of impossibility to approach to zero as close as necessary. Therefore, we match at certain point $x_{gl}$  the right profile  edge  with the linear approximation of solution located in the plane tangent to $W^s_{loc}$ at the origin. This trick results is as follows. Let $ Y=( v, u,w)$, then we compose the following profile   
$$Y=\left\{  
\begin{array}{cc}
Y_s,& x<x_{gl}, \\Y_s(x_{gl})\exp(\mbox{Re}(\lambda_2)(x-x_{gl})), &x>x_{gl}.
\end{array}\right.
$$
In fact, the position of gluing does not play an essential role as it will be shown below.  

Among the numerical methods \cite{Miura1982,Olmos2009,Carter2015} suitable for  simulation of  the wave movement, the method of lines \cite{methodLines} is used. Consider the spatial interval $x\in [a,b]$. 
 Let $x_n=a+h (n-1)$, $n=1,\dots K+1$, $h=(b-a)/K$, be the equidistant mesh points in $[a;b]$ and on these lines $V_n(t)=v(t,x_n)$, $W_n(t)=w(t,x_n)$. Then the numerical scheme is as follows 
\begin{equation*}
\begin{split}
\tau V_n  '' +V_n '&=(V_{n+1}-2V_n+V_{n-1})/h^2+f(V_n)-W_n, \\ W_n'&=\epsilon(V_n-\gamma W_n), \qquad n=2\dots K, 
\end{split}
\end{equation*}
subjected to the initial condition $V_n(0)=v_s(x_n)$,  $V_n'(0)=c u_s(x_n)$,  $W_n(0)=w_s(x_n)$. For the boundary points we prescribe $V_1(t)=v_s(a)$,  $V_{K+1}(t)=v_s(b)$,  $W_1(t)=w_s(a)$,  $W_{K+1}(t)=w_s(b)$.

Note that the result of application of method of lines coincides also with the using of ``Mathematica'' package command \textsf{NDSolve[$\,$]}. 
   
\begin{figure}
\centering
\subfloat[]{
\resizebox*{6cm}{!}{\includegraphics{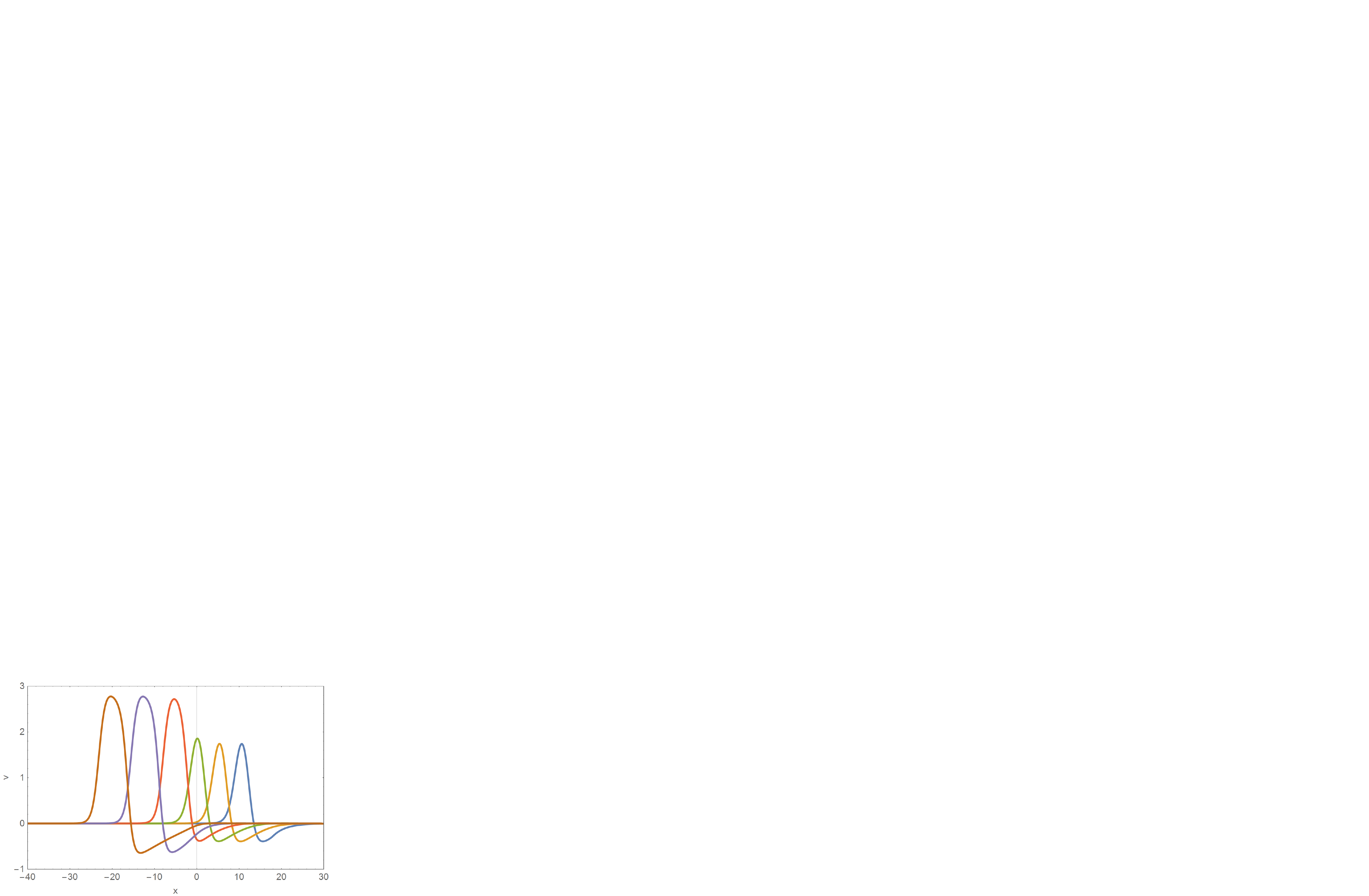}}}\hspace{5pt}
\subfloat[]{\resizebox*{6cm}{!}{\includegraphics{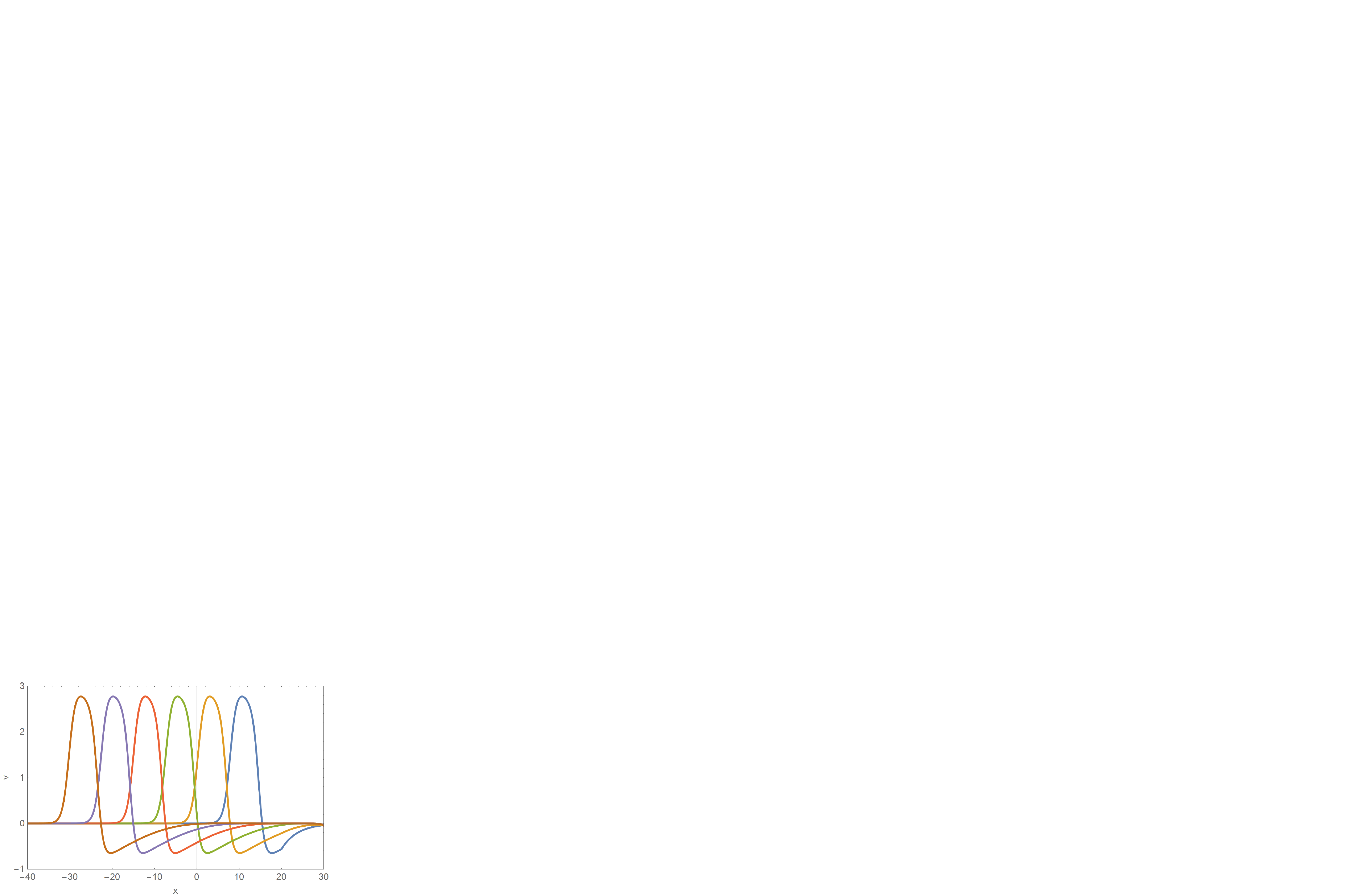}}}
\caption{The simulation of  slow (a) and fast (b) solitary waves' propagation at $\tau=0.2$. The profiles are drawn every 10 time units starting from $t=0$. The initial profiles are represented by the rightmost curves.} \label{vgs:fig3}
\end{figure}

    It follows from Figure~\ref{vgs:fig3}(a) that for some time the slow wave moves in a self-similar mode. Moreover, the weak nonsmoothness of the tail disappears. For $t>t^\star\approx 15$ the wave loses its stability, which is manifested in the deformation  of initial profile. It turns out that the height  of the wave grows causing the wave transformation  into the fast solitary wave, as it can be seen comparing the  resulting profiles in Figure~\ref{vgs:fig3}(a) and Figure~\ref{vgs:fig3}(b).
   
   The fast solitary wave shows essentially different behavior. According to Figure~\ref{vgs:fig3}(b), the wave propagates without change of the form, so it seems to be stable.  
   To be convinced that the above conjectures are valid, in the following section the solitary wave stability is studied via the Evans function technique \cite{Smereka}. 

\section{Study of spectral stability  of the solitary waves}\label{Sec:three}

In present studies we are going to deal with the spectral stability concept when the spectrum of linearized operator allows one to predict the behavior of small disturbances in a vicinity of specified solution.

The  operator in question is defined by the linearization of the system near the solitary wave solution $Y_s(z)=\left(v(z),\,u(z),\,w(z)\right)^{tr}$. In studying spectral stability it is instructive to pass in (\ref{PDEq}) to traveling wave coordinates $z=x+ct$, $\theta =t$ in which the solitary wave solution $Y_s(z)$ becomes stationary.  After doing this, we insert into  (\ref{PDEq}) the ansatz of special form
\begin{equation}\label{ans}
Y(z,\theta)=Y_s(z)+e^{\lambda\theta}y(z),
\end{equation}
where $y(z)=\left(y_1(z),\,y_2(z),\,y_3(z)\right)^{tr}$ is a small perturbation. In accordance with (\ref{VGS:uklad2}), we assume that $y_2(z)=d\,y_1(z)/d\,z.$ Inserting the ansatz (\ref{ans}) into (\ref{PDEq}), and dropping out the terms of the order $O(|y|^2,$  we get:
\[
\begin{array}{c}
y_1^\prime=y_2, \\
\tau\left(\lambda^2 y_1+2\,c\,y_2+c^2\,y_2^\prime   \right)+\lambda y_1+c\,y_2=y_2^\prime+M(z) y_1-y_3, \\
\lambda y_3+c\,y_3^\prime=\epsilon \left(y_1-\gamma y_3   \right),
\end{array}
\] 
where $M(z)=df(v(z))/dv$.
The above equations can be rewritten in the form of the following generalized eigenvalue problem:
 \begin{equation}\label{gvs:lins}
  \frac{dy}{dz}=\left(
 \begin{array}{ccc}
 0&1&0\\
 \{\tau \lambda^2+\lambda-M(z)\}\beta  & c(1+2\tau\lambda)\beta &\beta\\
 \epsilon/c & 0 &(-\epsilon\gamma -\lambda  )/c\\
   \end{array}
 \right) y =A(\lambda; Y_s(z))y.
 \end{equation}
Let us recall that if $\lambda$ belongs to the spectrum of the operator $L$, then the corresponding bounded solution $y$ satisfying specified boundary conditions is the eigenfunction corresponding to this eigenvalue.
 The wave $Y_s$ is said to be spectrally stable if the spectrum $\sigma(L)$ has no intersection with the positive half-plane of the complex plane, \cite{Sandstede, Blank2011}.

Investigation of the presence (or absence) of isolated eigenvalues ​​of the operator $L$ in the positive half-plane is performed with the help of the Evans function technique \cite{Smereka,Lega2011,Blank2011}. Within this technique, we are interested in the  set $Z^+ $ composed of solutions  of the
linear system (\ref{gvs:lins}) bounded as $z$ tends to $+\infty$ and the set $Z^-$ composed of solutions  bounded at $-\infty$. In case when the sets $Z^+ $ and $Z^- $ have a nonzero intersection, the linearized system has a bounded solution, and the eigenvalue for which this property takes place belongs to the discrete spectrum of the operator $L$.
The above condition is fulfilled, provided that the Wronskian of these  solutions nullifies. Thus,
the Evans function is defined as follows
$
E(\lambda)=\mbox{ det}(Z^-,Z^+)\Bigr|_{z=0}$.

The alternative definition can be formulated in the term of  transmission coefficient \cite{Yanagida,Pego1997,Smereka}. In paper \cite{Smereka} it has been considered the  case when the spectrum of $A(\lambda;Y_s(+\infty))$ contains only one eigenvalue $\mu_+$ with negative real part
, and   $A(\lambda;Y_s(-\infty))$ posseses only one eigenvalue $\mu_-$ with negative real part 
But $Z^+$, which is bounded at $+\infty$, as a solution of autonomous linear system possesses the component $E(\lambda)\eta_- e^{\mu_- x}$  at $-\infty$. Thus,  the bounded solution exists if the coefficient $E(\lambda)$ nullifies. In this context nullifying of the Evans function means the absence of unstable components in the set of solutions.  


When the homoclinic orbit is studied, as in our case, $A(\lambda;Y_s(-\infty))=A(\lambda;Y_s(+\infty)) = A(\lambda;0)\equiv A$. 
It can be checked, that for $\lambda>0$  the eigenvalues of the matrix $A^{\infty}$ satisfy the inequalities  $Re(\mu_{2,3})<0<Re(\mu_1)$. In order to get the appropriate signs of eigenvalues, the  temporal variable $z$ in the system (\ref{gvs:lins})  is replaced with the  variable $(-z)$ having the opposite sign. 

Applying this trick we arrive at the system whose eigenvalues satisfy the inequalities $Re(\mu_1)<0<Re(\mu_{2,3}).$ Then at $+\infty $ the system possesses one-dimensional stable component, while at $-\infty$ one-dimensional unstable and two-dimensional stable linear manifolds. If we start to integrate from stable solution at $+\infty$, then  we intersect the one-dimensional unstable component at $-\infty,$ defining the Evans function. This strategy is realized numerically.

It is  instructive to make the change of variables $y=e^{\mu_1 z}\bar y$ leading to the following representation: 
\begin{equation}\label{vgs:evans}
\bar y'=(-A(\lambda, Y_s(-z))-\mu_1 I) \bar y.
\end{equation}
We are looking  for the solution such that $\bar y,$ having the asymptotics $q=\{1;\mu_1/A_{12},A_{31}/(\mu_1-A_{33})\}$ as $z \rightarrow +\infty$, and $\bar y\rightarrow E(\lambda) q $ as $z \rightarrow -\infty$.
This system should be accompanied by the system describing the wave profile subjected by the proper initial condition. 
The solitary wave profile $Y_s$ is taken from Figure~\ref{vgs:fig3} when the initial profile has moved during 10 time units and got rid of  its original nonsmoothness. By placing this profile at the interval $z\in [0;40]$ we identify the $z_{+\infty}=40$ and $z_{-\infty}=0$.  
Thus, to get the Evans function values we should integrate (\ref{vgs:evans}) from $z_{+\infty}$ to $z_{-\infty}$ and extract the first coordinate of solutions $\bar y_1$. But to improve the convergence of $E(\lambda)$ evaluation, the recommendation from \cite{Smereka} can be applied and then $E(\lambda)=\bar y (z_{-\infty}) \cdot r$, where $r=\{(\mu_1-A_{22})/A_{12}; 1; A_{23}/(\mu_1-A_{33})\}/\Delta$ is the left eigenvector of $A$, $\Delta=(2\mu_1-A_{22})/A_{21}+(A_{31} A_{23})/(\mu_1-A_{33})^2$.

\begin{figure}
\centering
\subfloat[]{
\resizebox*{4cm}{!}{\includegraphics{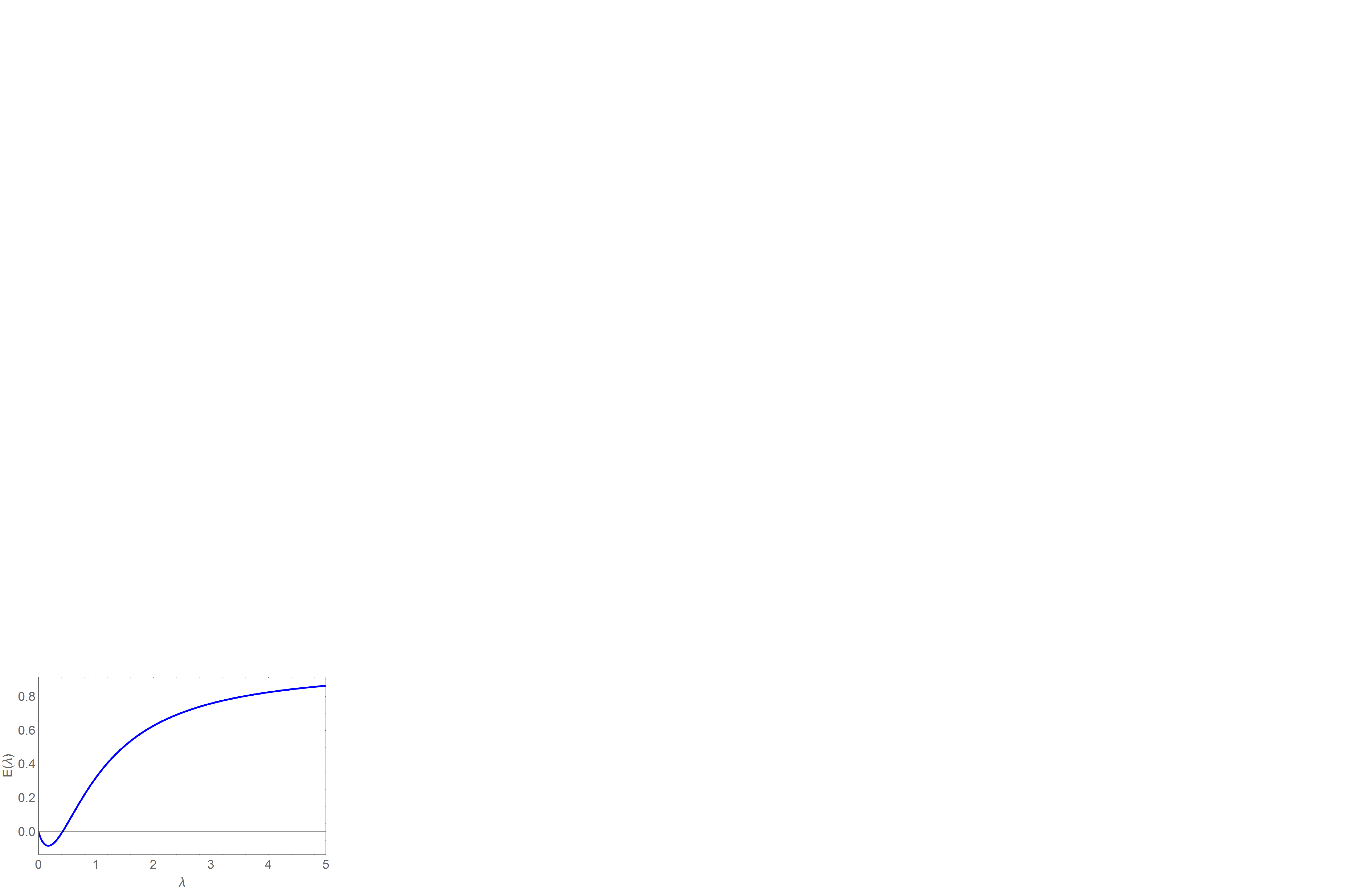}}}\hspace{5pt}
\subfloat[]{\resizebox*{4cm}{!}{\includegraphics{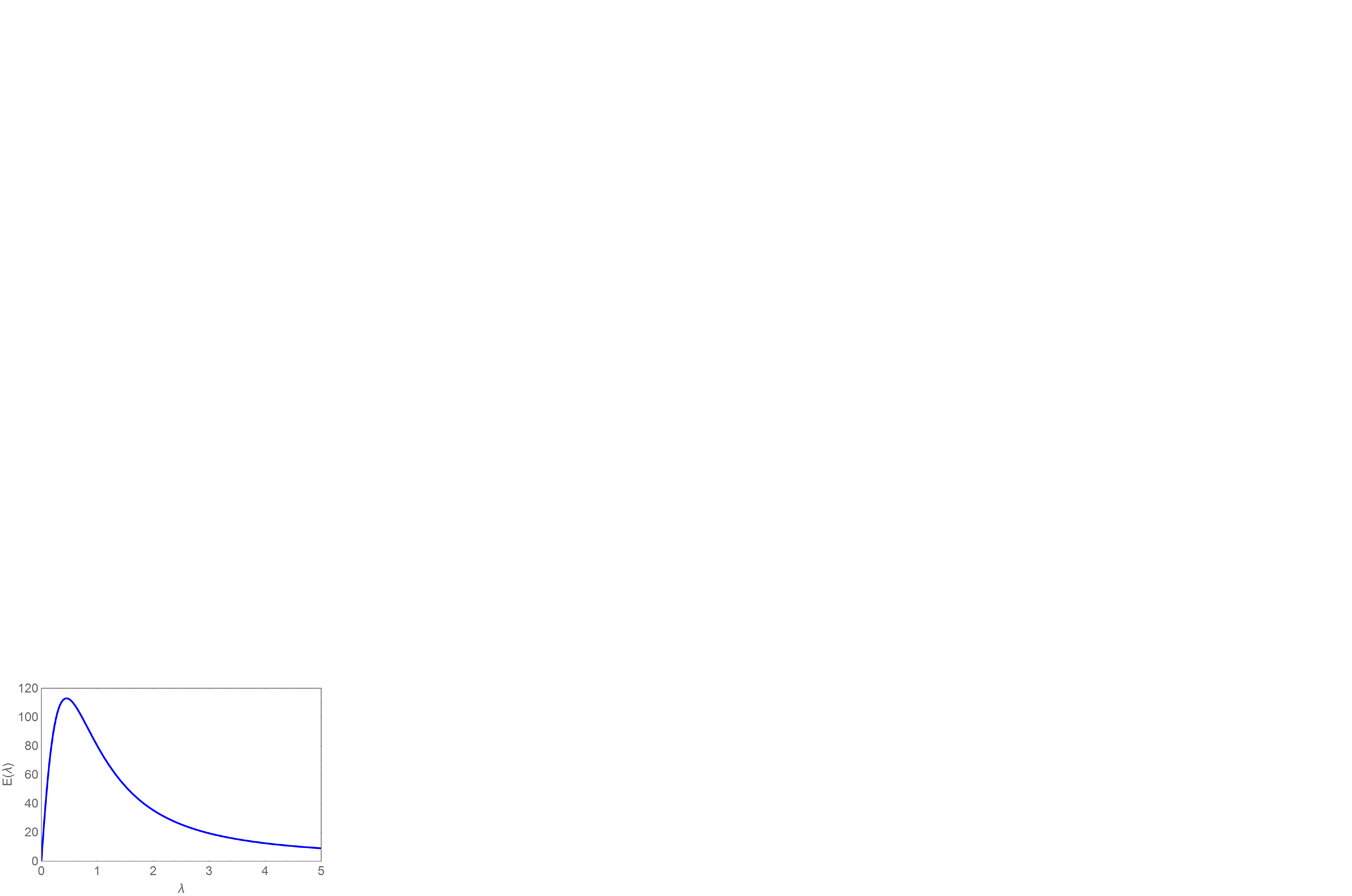}}}
\caption{The graphs of Evans function for the  fast (a) and slow (b) solitary waves corresponding to the initial profiles in Figure~\ref{vgs:fig3}.  } \label{vgs:fig4}
\end{figure}

At first, using the procedure outlined, the real valued Evans function $E(\lambda)$ 
is evaluated.  Figure~\ref{vgs:fig4}(a)  illustrates the graph of  $E(\lambda)$ corresponding to the slow solitary wave shown in Figure~\ref{vgs:fig3}(a), whereas Figure~\ref{vgs:fig4}(b) relates to the fast solitary wave depicted in  Figure~\ref{vgs:fig3}(b). Since in the former case $E(\lambda)$ possesses the positive root, then the slow solitary wave is unstable that agrees with the simulation results. The latter case tells us that there are no positive real roots only and thus we should consider in addition the possibility of  complex roots existence.   

To do this,  the Nyquist diagram technique \cite{Lega2011,Blank2011} can be used. This method allows one to estimate a number of zeros located  in the closed region for an analytic function. More precisely, the difference between numbers of zeros and poles is equal to a number of  encircling the origin by the curve as it is prescribed by  Cauchy's argument principle. 

\begin{figure}
\centering
\subfloat[]{
\resizebox{!}{5cm}{\includegraphics{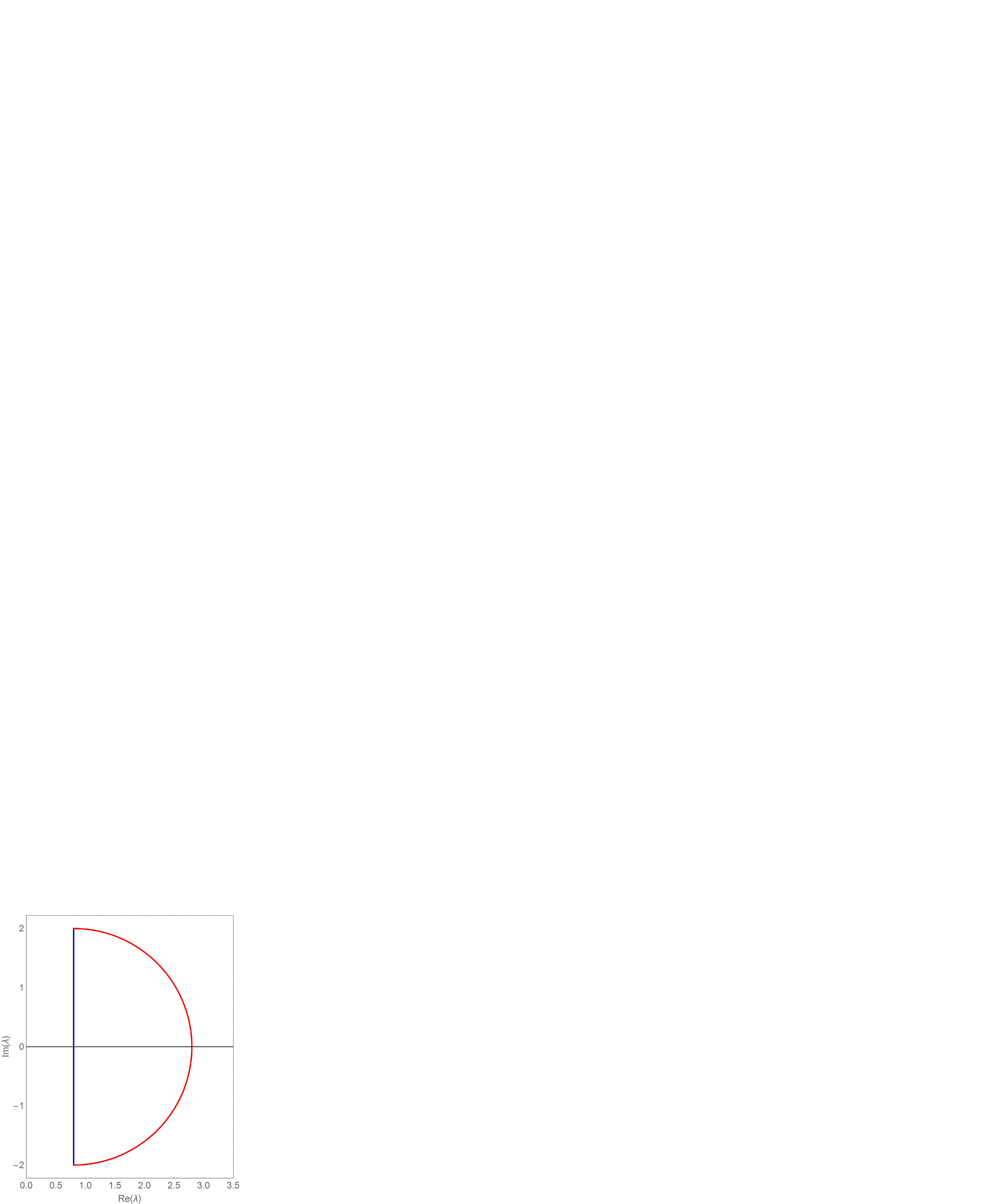}}}\hspace{5pt}
\subfloat[]{\resizebox{!}{5cm}{\includegraphics{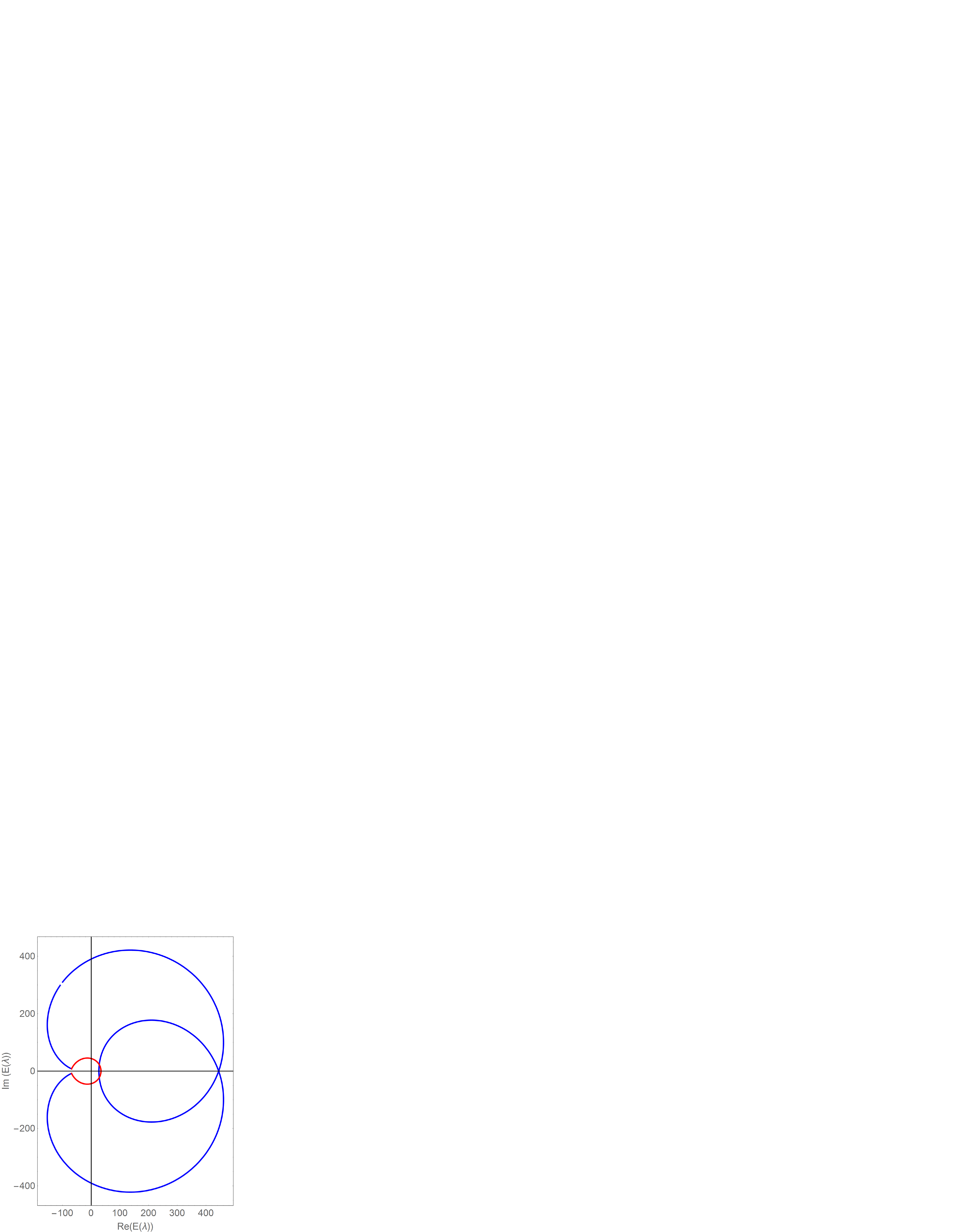}}}
\caption{The contour (a) and the corresponding Nyquist diagram (b)  for the Evans function of the fast solitary wave. } \label{vgs:fig51}
\end{figure}

Let us choose the contour in the form of half-circle with radius equal 2, see  Figure~\ref{vgs:fig51}(a), then their image via the Evans function is shown in Figure~\ref{vgs:fig51}(b).

As we can see, the curve in Figure~\ref{vgs:fig51}(b) does not surround the origin. Therefore, there are no zeros inside the contour shown in Figure~\ref{vgs:fig51}(a). The same result is obtained when taking the  contours of larger radius. Thus, the application of the Evans function technique confirms the  instability of slow wave and does not reveal the  instability of fast one.

\section{On the formation of solitary waves and their behavior during mutual collisions}\label{Sec:four}

{\it Piston problem for mFHN equation}

Motivated by the question on the possible source of solitary wave appearance, we have opted to solve the piston problem for the mFHN model. 

To do this, the appropriate boundary conditions should be prescribed. We considered the interval $x\in [0,L]$, where  $L=100$ is large enough  to make it possible the observation of the localized  modes formation. Next, we fixed the initial condition as follows $v(t=0,x)=0$, $w(t=0,x)=0$. The piston at left edge had the velocity $U_0$ and moved to the right.  So, the boundary condition   $v_t(t,x=0)=U_0$, $v(t,x=L)=0$   (or $v_x(t,x=L)=0$) were posed. To match the velocity value at left edge and its value at $t=0$, the initial profile for $v_t$ was chosen in the form $
v_t=U_0\,\delta(x).$

\begin{figure}
\centering
\subfloat[]{
\resizebox*{!}{4cm}{\includegraphics{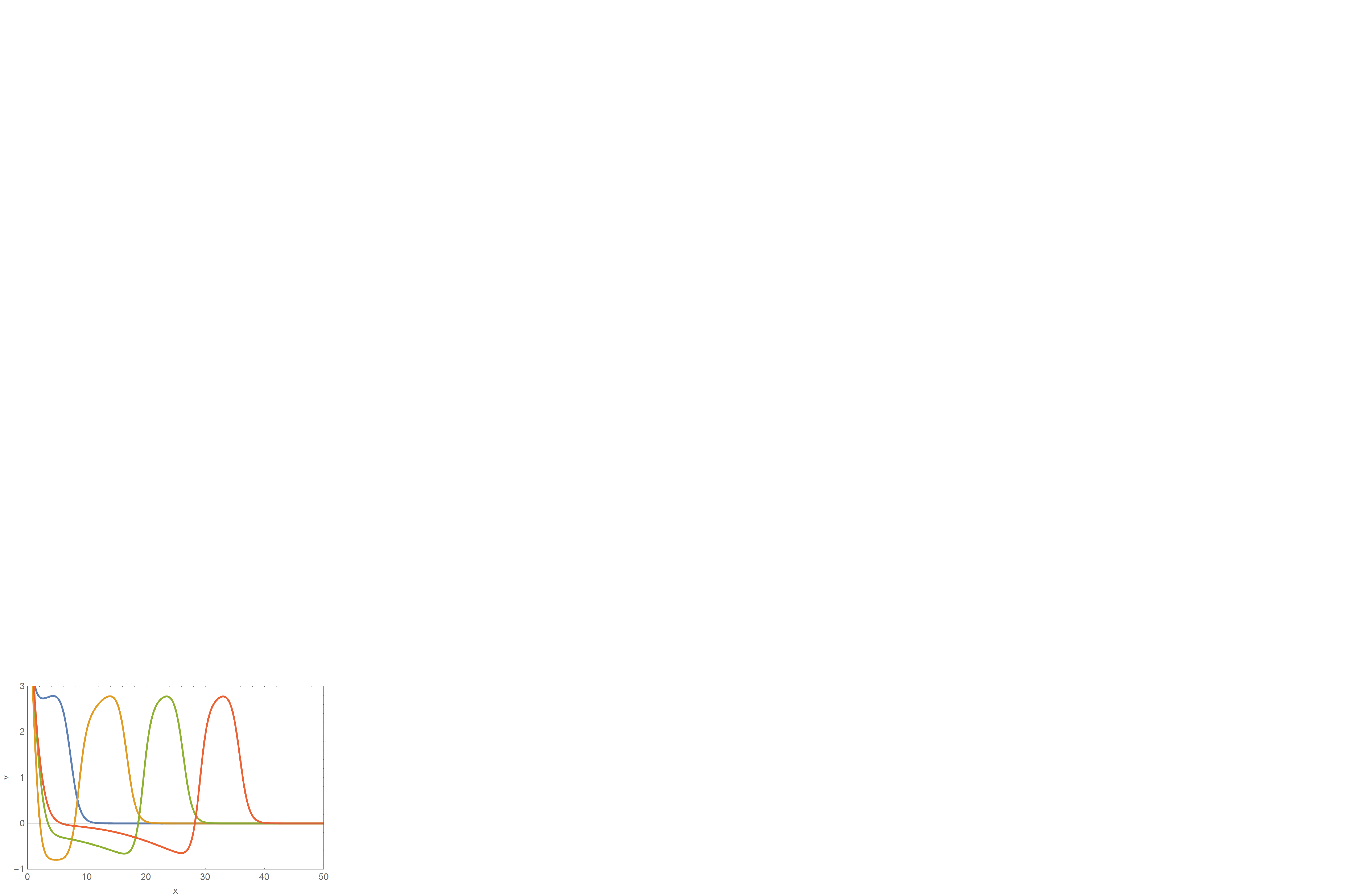}}}\hspace{5pt}
\subfloat[]{\resizebox*{!}{4cm}{\includegraphics{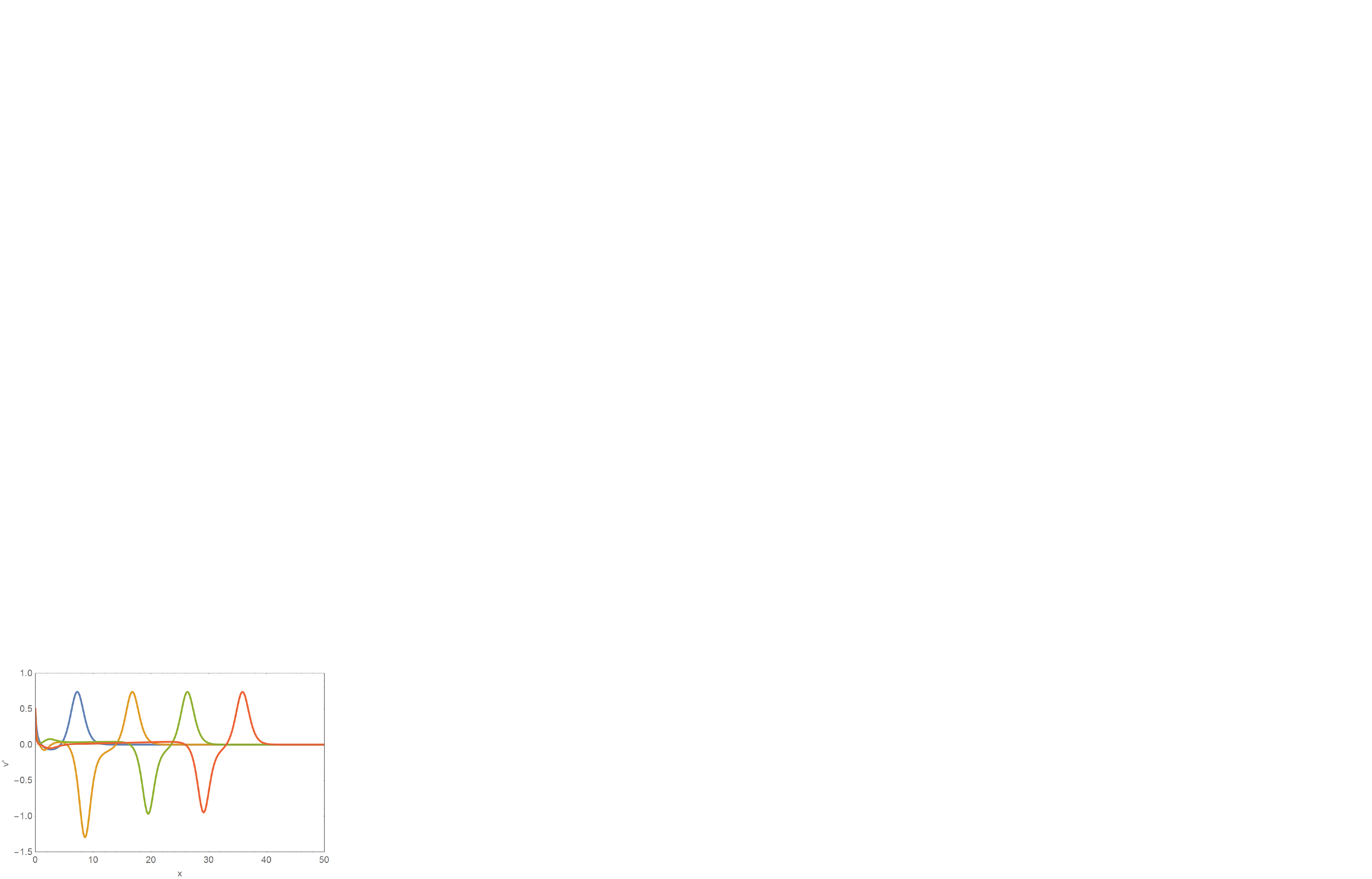}}}
\caption{The $v(t,x)$ (a) and $v_t(t,x)$ (b) components of  the solution to the piston problem. The initial piston velocity $U_0=0.5$. The total time of evolution $T=50$, the drawn profiles at $Tk/4$, $k=1,2,3,4$. } \label{vgs:fig6}
\end{figure}

The results of simulation for $U_0=0.5$ is presented in Figure~\ref{vgs:fig6}. We see that after some period of time  (depending on $\tau$) the profile resembling the fast solitary wave is forming. Comparing the final profiles in Figure~\ref{vgs:fig6} and profiles in Figure~\ref{vgs:fig1}(c) we can conclude that this is indeed the fast solitary wave  observed at the end of simulation. 

For $U_0<0.1$  the solitary wave formation was never observed. Thus, there is the threshold of $U_0$ separating the values at which in front of piston the solitary wave can be formed and values when the localized regimes do not appear.

{\it Solitary waves collision}

Another interesting problem is the solitary wave behavior  during the mutual collisions. According to the Figure~\ref{vgs:fig6}, the identical fast  solitary waves moving towards each other annihilate  after the collision. At the beginning they combine and form a one-humped standing wave, see Figure~\ref{vgs:fig6}(a), which is destroyed as the time passes by, see Figure~\ref{vgs:fig6}(b),(c).
\begin{figure}
\centering
\subfloat[]{
\resizebox*{4cm}{!}{\includegraphics{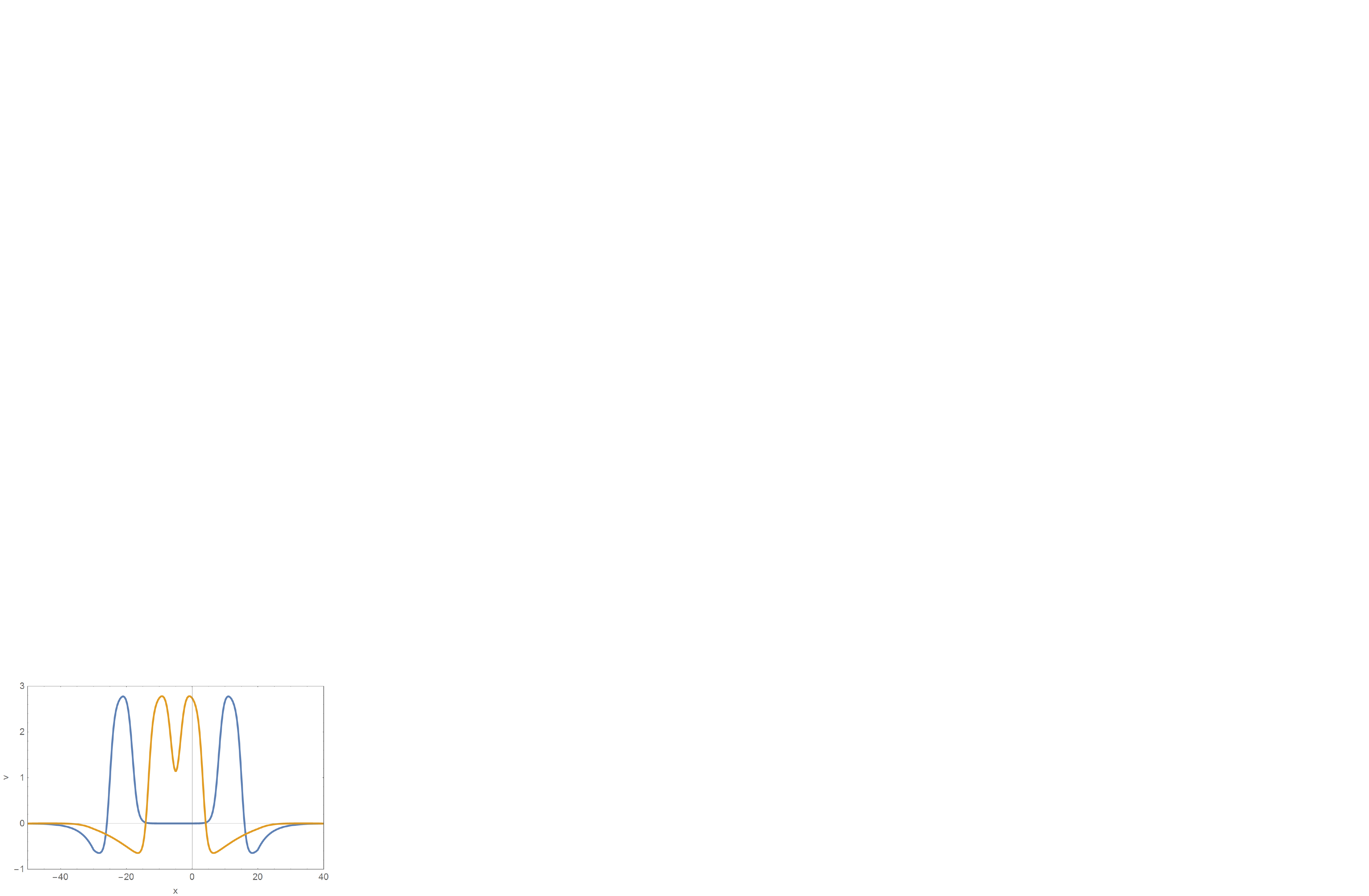}}}
\hspace{3pt}
\subfloat[]{
\resizebox*{4cm}{!}{\includegraphics{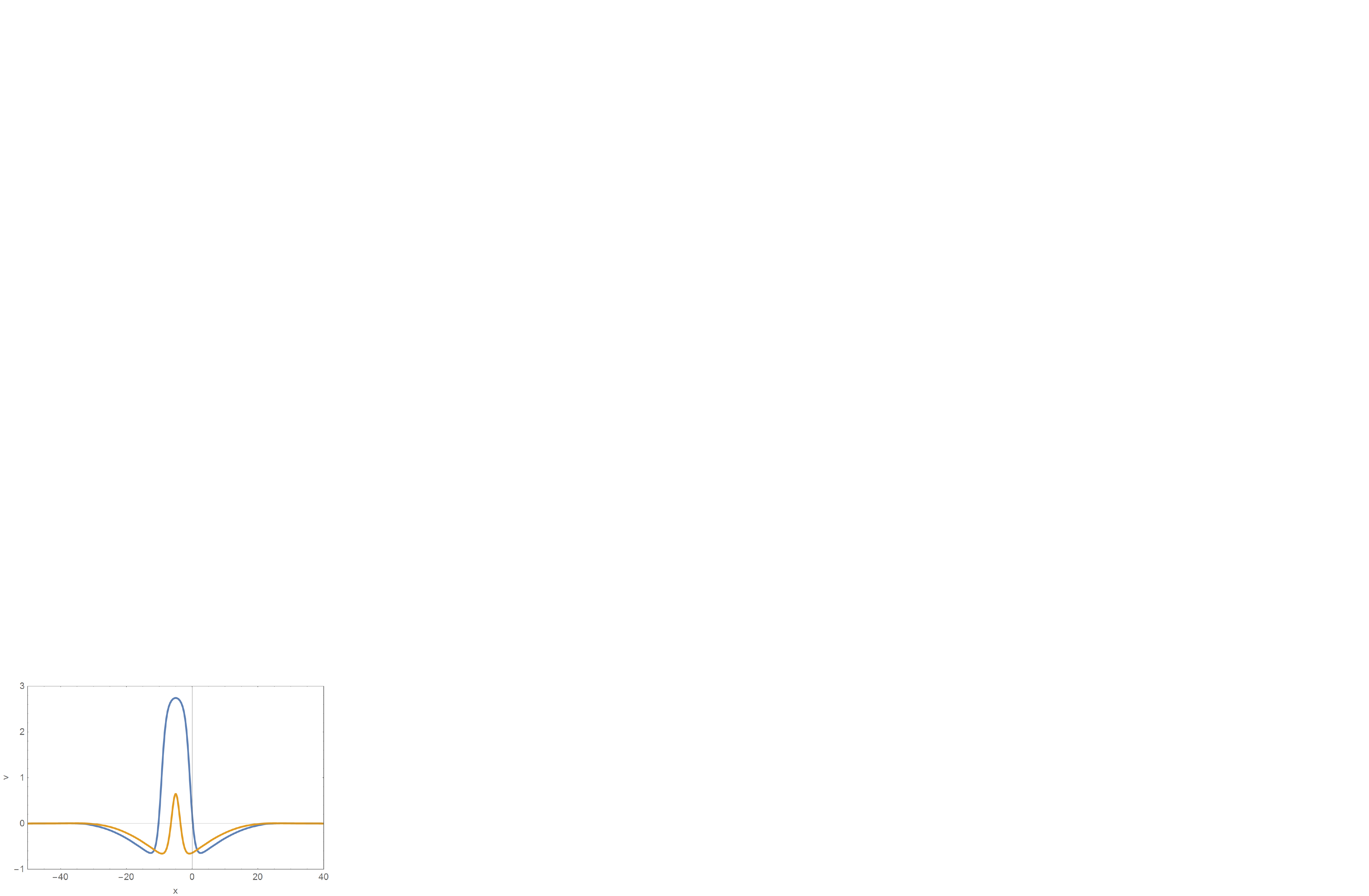}}}
\hspace{3pt}
\subfloat[]{
\resizebox*{4cm}{!}{\includegraphics{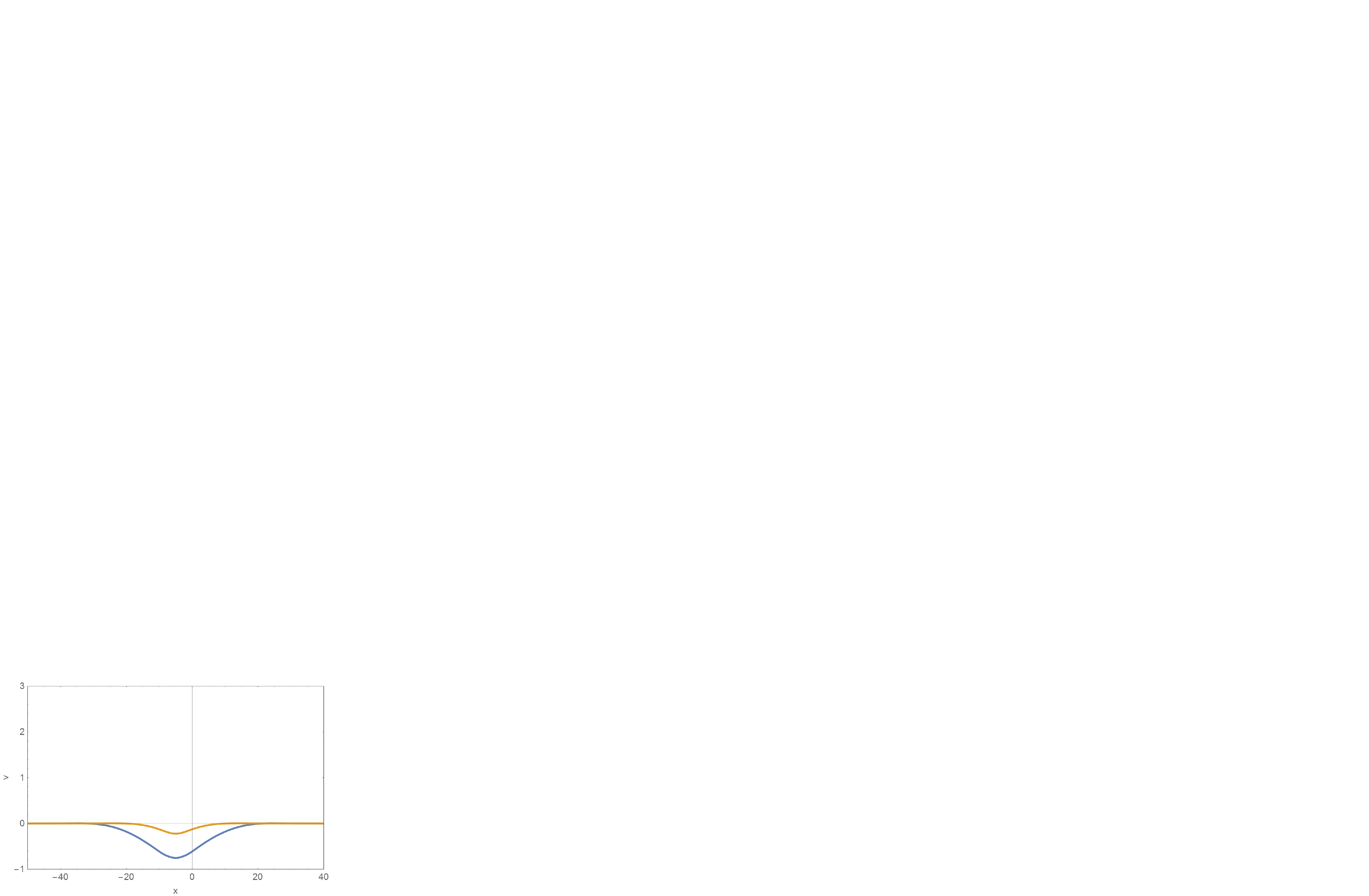}}}
\caption{The snapshots of fast solitary waves interaction for $t=0$, $t=15$ on graph (a), $t=20$, $t=24$ on (b) and $t=25$, $t=40$ on (c).} \label{vgs:fig7}
\end{figure}

\section{Concluding remarks}\label{Sec:concl}

To sum up, we have shown that, like the classical FHN model, the modified  FitzHugh-Nagumo system (\ref{PDEq}) possesses  the solitary waves solutions. At fixed parameters $\epsilon, \gamma$ the solitary waves solutions are observed along the "banana-shaped" curves in  the parameter space $(a,\,c).$ In particular, the larger is the value of $\tau$, the more lean the banana-shaped curve becomes. Increasing of $\tau$ also causes   the decreasing of maximal solitary wave velocity, which is not surprising due to the inequality $\beta>0.$ 

Simulations of the single wave movement reveal that the fast solitary wave propagates  in a self-similar mode, whereas the slow solitary wave with time begins to grow and finally transforms into the fast solitary wave. The latter means the instability of slow pulses. The simulation  findings concerning the stability  property of waves are confirmed by the analysis of the Evans function. It turns out that there is the positive real root of the Evans function evaluated for the slow pulse, which proves its instability. In the case of the fast pulse no real or complex Evans function roots in the right complex half-plane are fixed.  

It is also shown that the fast solitary waves can appear in front of  the piston moving  into the medium described by the mFHN equations provided that its velocity exceeds some threshold value. The annihilation of two fast (stable) pulses of the same forms during the mutual is observed. The latter is rather typical for the dissipative systems.

\section*{Acknowledgements}

The investigations carried out by two authors (AG and VV) were partially supported
by the Faculty of Applied Mathematics AGH UST within subsidy of Ministry of Science
and Higher Education of Poland, grant \# 11.11.420.004

\end{document}